\documentclass[conference]{IEEEtran}
\IEEEoverridecommandlockouts
\usepackage{amsmath}
\usepackage{amsfonts}
\usepackage{relsize}
\usepackage{tikz}
\usepackage{algorithm}
\usepackage{algorithmicx}
\usepackage{algpseudocode}
\usepackage{mathtools}
\usepackage{dsfont}
\usepackage{accents} 
\usepackage{float}
\usepackage{subcaption}
\usepackage{cuted}
\usepackage{flushend}
\usepackage{stfloats}
\usepackage{comment}
\usepackage{xcolor}
\usepackage{colortbl}
\usepackage{hhline}

\newcommand{\vect}[1]{\boldsymbol{#1}}

\DeclareMathOperator{\argmax}{argmax}

\definecolor{darkgreen}{rgb}{0,0.75,0}
\definecolor{maroon}{RGB}{186,0,0}
\definecolor{purple}{RGB}{96,26,149}
\definecolor{mavi}{RGB}{46,76,255}
\definecolor{haki}{RGB}{38,99,33}

\begin{document}
\title{RIS-Assisted Grant-Free NOMA}

\author{\IEEEauthorblockN{Recep Akif Tasci\textsuperscript{$\ast$}, Fatih Kilinc\textsuperscript{$\ast$}, Abdulkadir Celik\textsuperscript{$\bullet$}, Asmaa Abdallah\textsuperscript{$\bullet$}, Ahmed M. Eltawil\textsuperscript{$\bullet$} and Ertugrul Basar\textsuperscript{$\ast$}}
\IEEEauthorblockA{\textsuperscript{$\ast$}Communications Research and Innovation Laboratory (CoreLab),\\Department of Electrical and Electronics Engineering, Ko\c{c} University, Sariyer 34450, Istanbul, Turkey. \\
\textsuperscript{$\bullet$}Computer, Electrical, and Mathematical Sciences \& Engineering (CEMSE) Division,\\ King Abdullah University of Science and Technology (KAUST),
Thuwal, KSA 23955-6900 \vspace{-1cm}
}
\thanks{The work of E. Basar was supported in part by TUBITAK under Grant 120E401. R. A. Tasci and F. Kilinc worked on this project during an internship at King Abdullah University of Science and Technology (KAUST).

}
}

\maketitle

% As a general rule, do not put math, special symbols or citations
% in the abstract
\begin{abstract}
This paper introduces a reconfigurable intelligent surface (RIS)-assisted grant-free non-orthogonal multiple access (GF-NOMA) scheme. To ensure the power reception disparity required by the power domain NOMA (PD-NOMA), we propose a joint user clustering and RIS assignment/alignment approach that maximizes the network sum rate by judiciously pairing user equipments (UEs) with distinct channel gains, assigning RISs to proper clusters, and aligning RIS phase shifts to the cluster members yielding the highest cluster sum rate. Once UEs are acknowledged with the cluster index, they are allowed to access their resource blocks (RBs) at any time requiring neither further grant acquisitions from the base station (BS) nor power control as all UEs are requested to transmit at the same power. In this way, the proposed approach performs an implicit over-the-air power control with minimal control signaling between the BS and UEs, which has shown to deliver up to 20\% higher network sum rate than benchmark GF-NOMA and grant-based optimal (OPT) PD-NOMA schemes depending on the network parameters. The given numerical results also investigate the impact of UE density, RIS deployment, and RIS hardware specifications on the overall performance of the proposed RIS-aided GF-NOMA scheme.  
\end{abstract}
\begin{IEEEkeywords}
Reconfigurable intelligent surface, power domain, grant-free, non-orthogonal multiple access, clustering, resource allocation.
\end{IEEEkeywords}
\IEEEpeerreviewmaketitle

\section{Introduction}
Reconfigurable intelligent surfaces (RISs) have emerged as a transformative technology envisioned for next-generation wireless networks to improve the spectral efficiency (SE) with low power consumption and hardware costs \cite{basar2019wireless}. RISs have low-cost and low-power hardware as they consist of a large number of passive elements. The RIS controller manipulates the reflection of incident electromagnetic waves towards the desired direction by intelligently controlling the phase shift of elements \cite{arslan2021over}. The passive beamforming nature of RISs has been shown to improve network capacity \cite{basar2019wireless,arslan2021over}, enhance the end-to-end performance of multi-hop communications \cite{kilinc2021physical}, and deliver a performance comparable to traditional relaying schemes \cite{marcorisrelay}. 

Moreover, non-orthogonal multiple access (NOMA) has also been recognized as a promising candidate to enable massive machine-type communication by multiplexing several user equipments (UEs) on the same network resources \cite{nomailk,nomapotentials}. In particular, power domain NOMA (PD-NOMA) improves the SE by multiplexing UEs with different channel gain and controlling transmit power such that successive interference cancellation (SIC) can be performed at the receiver. As the network size increases, PD-NOMA starts suffering from power control complexity, user pairing and resource allocation overhead, and ramifications of channel state information (CSI) acquisition \cite{gfnomasurvey}. NOMA has also been investigated with emerging technologies such as index modulation \cite{arslannoma}, cognitive radios \cite{sultancr}, and unmanned aerial vehicles-aided wireless networks \cite{sultanuav}. Recently, RISs have received attention thanks to their ability to create the reception power disparity required by PD-NOMA. While the RIS-aided NOMA is analyzed in \cite{RisnomaHou}, dynamic and static RIS configurations for multi-user NOMA schemes are considered in \cite{liu2020reconfigurable}. The authors of \cite{aymennoma} studied partitioning of RIS to facilitate PD-NOMA. Despite the fact that all of the aforementioned systems are grant-based (GB), there is still an urgent need to address the signaling overhead and computational complexity issues that are a bottleneck for NOMA technology. This is why there has lately been a lot of interest in the study of grant-free NOMA (GF-NOMA) schemes that UEs can operate without grant acquisition. The purpose of the PD GF-NOMA scheme is to eliminate the need of the complex power control operation, where the UEs can instead use fixed powers and rely on the channel gains disparity which leads to implicit over-the-air power control and reception power disparity. 
% However, none of these works consider the GF-NOMA scheme, which is necessary to make NOMA feasible for massive connectivity \cite{celik2022gfnoma}. 
Multiple access based \cite{gfnomasurvey}, compute and forward based, compressing sensing based \cite{LiuStochasticGFNOMA}, index modulation based \cite{DoganGFNOMA}, and the transmit power pool based \cite{FayazGFNOMA} GF-NOMA schemes are studied in the literature. ALOHA-NOMA schemes \cite{BaleviALOHANOMA} and semi GF-NOMA scheme where the GB and GF users share the same spectrum is proposed in \cite{ZhangStochasticGFNOMA}. As pointed out by a recent survey on GF-NOMA, RIS-aided GF-NOMA is a promising technology but has not been explored in-depth yet \cite{gfnomasurvey}. To the best of the authors' knowledge, RIS-aided GF-NOMA is considered only in \cite{dprisgf}, where authors propose a semi GF-NOMA by using deep reinforcement learning to control power and phase shifts.

In this paper, we propose an RIS-aided GF-NOMA scheme based on a novel joint user clustering and RIS assignment/alignment mechanism. Once UEs are acknowledged with a cluster index, they are allowed to access their RBs at any time requiring neither further grant acquisitions from the BS nor power control as all UEs are requested to transmit at the same power. The required power disparity is obtained by three levels of implicit power control: 1) by clustering UEs with different channel gains into the same cluster, 2) properly assigning RISs to clusters to increase power reception disparity, and 3) aligning the phase shift matrix of RIS to the cluster members giving the highest possible cluster sum rate. In this way, the proposed approach performs over-the-air power control by judiciously pairing UEs, assigning RISs, and aligning phase shifts of RISs. Since this joint combinatorial problem is known to be NP-Hard, we develop an iterative 3D assignment approach that forms clusters and assigns/aligns RISs to maximize the network sum rate. The proposed scheme has shown to outperform GB optimal (OPT) PD-NOMA and GF-NOMA schemes.

The rest of this paper can be summarized as follows. In Section \ref{sec:sys}, we present the considered network and channel model, the proposed signal model and the RIS phase alignment process. In Section \ref{sec:def}, we discuss the problem formulation and explain the proposed solution methodology, and present the proposed iterative 3D clustering approach in Section \ref{sec:Iterative}. Finally, in Section \ref{sec:numresult}, we present numerical results and the paper is concluded in Section \ref{sec:conc}.

\section{Network Model}
\label{sec:sys}
We consider uplink (UL) operation of a network consisting of a single BS serving $U$ UEs over $R$ RBs with $W$ [Hz] bandwidth, whose sets are denoted by $\mathcal{U}$ and $\mathcal{R}$, respectively. All UEs operate at an identical power, $p_\text{id}$, which is broadcast by the BS over a control channel. Without loss of generality, UEs are assumed to be uniformly distributed over a cell area of radius $D$. To show proposed GF-NOMA schemes suitability for massive connectivity, we consider a dense network scenario (i.e., $U \gg R$) and group multiple UEs to utilize a single resource block. That is, the number of clusters $C$ is the same with $R$, and RB/cluster terms are used interchangeably throughout the paper. Even though $R$ and $C$ are fixed, $U$ can dynamically change according to spatio-temporal characteristics of the network traffic. Hence, the maximum cluster size is set as $K=\lceil \frac{U}{R} \rceil$. Accordingly, the cluster set utilizing RB$_r$ is defined by $\mathcal{C}_r=\{\textit{UE}_u \: \vert  \:  \chi_r^u=1, \forall u \in \mathcal{U}, \sum_{u \in \mathcal{U}} \chi_r^u \leq K\}$, where $\chi_r^u$ is the binary indicator, $\chi_r^u=1$ if $\textit{UE}_u$ belongs to $r$th cluster, otherwise $\chi_r^u=0$. 

\begin{figure}[t]
    \centering
    \includegraphics[width=0.85\columnwidth]{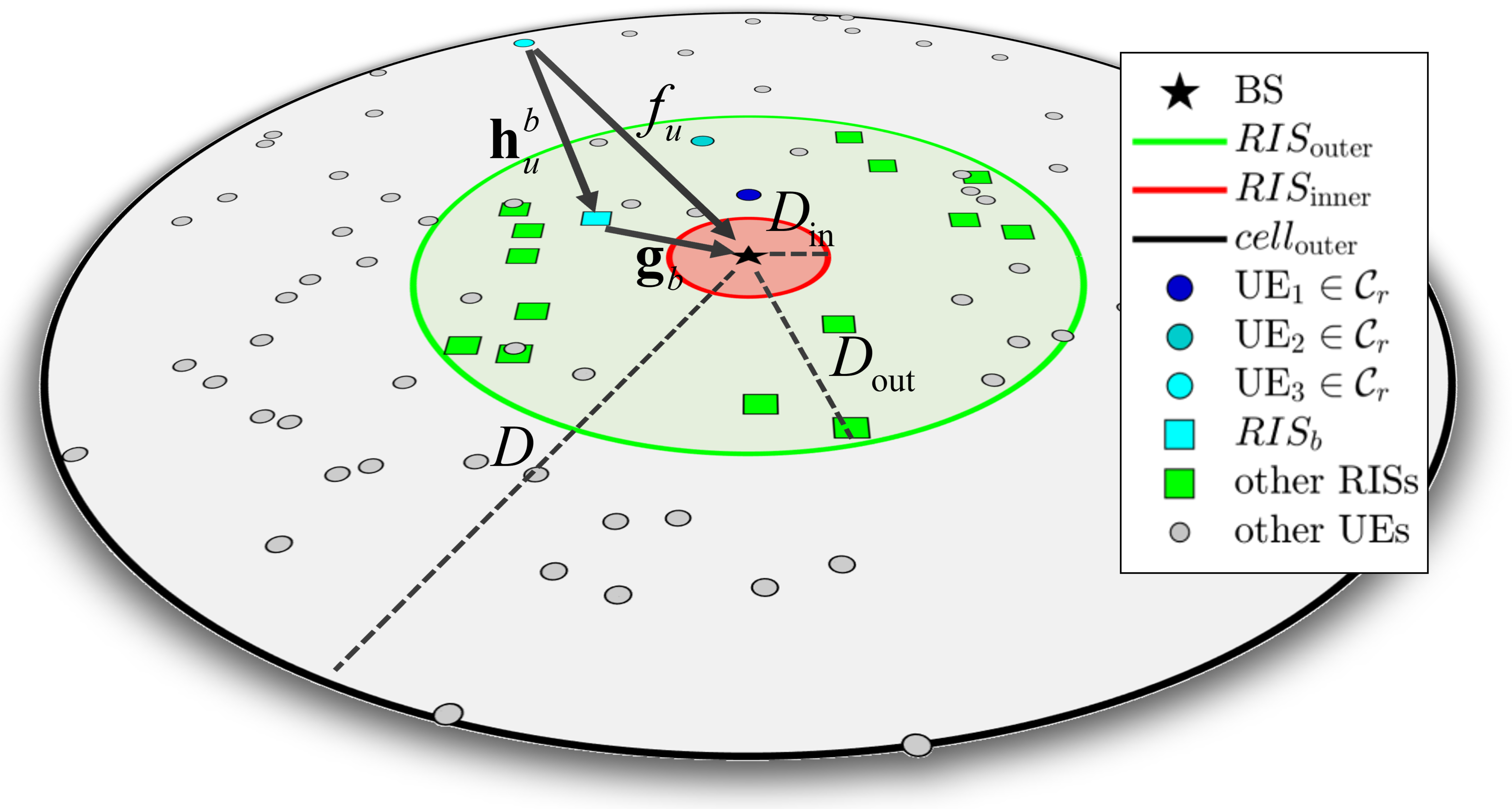}
    \vspace{0.15cm}
    \caption{Network model}
    \label{fig:ntwmodel}\vspace*{-0.30cm}
\end{figure}

In order to improve network performance and facilitate the GF-NOMA, $M$ RISs are uniformly deployed in a disk having inner and outer radius of $D_\text{in}$ and $D_\text{out}$, where $D_\text{in}$ ensures that RISs are placed at the far-field of the BS as shown in Fig. \ref{fig:ntwmodel} and $D_\text{in} \leq D_\text{out} \leq D$. Each physical RIS is partitioned into $G$ logical sub-blocks, each with $N$ elements, resulting in an overall $B=MG$ RIS blocks. Each block can be exploited by a member of a cluster, and a cluster cannot be assigned to more than one block. Accordingly, the set of RIS blocks are denoted by $\mathcal{B}$, and the $b$th RIS block is denoted by RIS$_b$. The binary RIS assignment matrix is denoted by $\boldsymbol{\Delta} \in \{0,1\}^{B \times U \times R}$ with entries $\delta_{r,b}^u$ such that $\delta_{r,b}^u=1$ if RIS$_b$ is assigned to the $r$th cluster and configured based on the channels of UE$_u \in \mathcal{C}_r$, $\delta_{r,b}^u=0$ otherwise. Fig. \ref{fig:ntwmodel} demonstrates the considered network model where the users in a cluster are shown with the circles in the shades of blue, the RIS assisting the cluster with the blue square, the base station with a black star, the other users in the network with gray circles and the other RISs with green squares assuming there are three users in a cluster. With the help of the 3D clustering algorithm, which will be explained in the next section, we will jointly form the RIS assignment matrix and assign users to clusters according to the modified received power of the users to maximize the sum rate.
% The transmit power of UE's can be adjusted according to the applied scheme and the tranmist power vector of the users in a cluster is denoted by $\mathbf{p}_t$. The maximum and minimum transmit powers available to a user are indicated by $P_{max}$ and $P_{min}$, respectively. The benchmark multi-level grant-free NOMA (MGF-NOMA) scheme  identifies the transmission power levels of the users by linearly dividing the $[P_{max}, P_{min}]$ into $K$ levels. If the benchmark single-level grant-free NOMA (SGF-NOMA) and RIS-assisted GF-NOMA is applied, all users transmit at the same power which is denoted by $P \in [P_{max},P_{min}] $.
% \subsection{Channel Model}
% \label{sec:ch}

The channels are modeled according to the Urban Macro (UMa) scenario of the 3GPP standard specified for 0.5-100 GHz bands \cite{3GPP_5G}. The channel between UE$_u$ and RIS$_b$ is denoted by $\mathbf{h}_u^b= \left[h_{u,b}^1,\ldots, h_{u,b}^n,\ldots,h_{u,b}^N\right]\in \mathbb{C}^{N\times 1}$, where $h_{u,b}^n$ is the channel coefficient from UE$_u$ to $n$th reflecting element of the RIS$_b$,  $n \in [1,N]$. Similarly, the channel vector between the BS and RIS$_b$ is denoted by $\mathbf{g}_{b}=\left[g_b^1,\ldots, g_b^n,\ldots,g_b^N\right] \in \mathbb{C}^{N\times 1} $,  where $g_b^n$ is the channel coefficient between the BS and $n$th reflecting element of the RIS$_b$, $n \in [1,N]$. Moreover, the direct channel from UE$_u$ to the BS is denoted by $f_u\in \mathbb{C}^{1\times 1}$. All these channels can experience Rayleigh and Rician fading under line-of-sight (LoS) and non-line-of-sight (NLoS) conditions, respectively. The channel coefficients are modeled as
% \begingroup
% \setlength\abovedisplayskip{10pt}
% \setlength\belowdisplayskip{10pt}
% \begin{align}\label{eq:h_coef}
%     h_{u,b}^n =  \!\sqrt{\frac{1}{\lambda_h}} \!\left(\!\sqrt{\frac{K_h}{K_h+1}} h_{u,b}^{n,\text{LoS}} + \sqrt{\frac{1}{K_h+1}}h_{u,b}^{n,\text{NLoS}}\right),
% \end{align}
% \endgroup
% \begingroup
% \setlength\abovedisplayskip{10pt}
% \setlength\belowdisplayskip{10pt}
% \begin{align}\label{eq:g_coef}
%     g_b^n =  \sqrt{\frac{1}{\lambda_g}} \left(\sqrt{\frac{K_g}{K_g+1}} g_b^{n,\text{LoS}} + \sqrt{\frac{1}{K_g+1}}g_b^{n,\text{NLoS}}\right),
% \end{align}
% \endgroup
% \begingroup
% \setlength\abovedisplayskip{10pt}
% \setlength\belowdisplayskip{10pt}
% \begin{align}\label{eq:f_coef}
%     f_u =  \sqrt{\frac{1}{\lambda_f}} \left(\sqrt{\frac{K_f}{K_f+1}} f_{u}^\text{LoS} + \sqrt{\frac{1}{K_f+1}}f_{u}^\text{NLoS}\right),
% \end{align}
% \endgroup
\begin{align}\label{eq:h_coef}
    h_{u,b}^n =  \!\sqrt{\frac{1}{\lambda_h}} \!\left(\!\sqrt{\frac{K_h}{K_h+1}} h_{u,b}^{n,\text{LoS}} + \sqrt{\frac{1}{K_h+1}}h_{u,b}^{n,\text{NLoS}}\right),
\end{align}
\begin{align}\label{eq:g_coef}
    g_b^n =  \sqrt{\frac{1}{\lambda_g}} \left(\sqrt{\frac{K_g}{K_g+1}} g_b^{n,\text{LoS}} + \sqrt{\frac{1}{K_g+1}}g_b^{n,\text{NLoS}}\right),
\end{align}
\begin{align}\label{eq:f_coef}
    f_u =  \sqrt{\frac{1}{\lambda_f}} \left(\sqrt{\frac{K_f}{K_f+1}} f_{u}^\text{LoS} + \sqrt{\frac{1}{K_f+1}}f_{u}^\text{NLoS}\right),
\end{align}
where $K_h$/$K_g$/$K_f$ is the Rician factor for 
$\mathbf{h}_u^b/\mathbf{g}_b/f_u$; $\lambda_h$/$\lambda_g$/$\lambda_f$ is the path loss over $\mathbf{h}_u^b/\mathbf{g}_b/f_u$; $h_{u,b}^{n,\text{LoS}}$/$g_b^{n,\text{LoS}}$/$f_{u}^\text{LoS}$ is the LoS component and $h_{u,b}^{n,\text{NLoS}}$/$g_b^{n,\text{NLoS}}$/ $f_{u}^\text{NLoS}\sim\mathcal{CN}(0,1)$ is the NLoS component of  $\mathbf{h}_u^b/\mathbf{g}_b/f_u$. If a channel does not include a LoS component, which mostly refers to Rayleigh fading, we consider the corresponding Rician factor as 0. $\lambda_h,\lambda_g$ and $\lambda_f$ are calculated according to the UMa scenario of the 3GPP standard considering the LoS probability specified in \cite{3GPP_5G}.

Based on UE-RIS-BS cascaded channels and UE-BS channels defined in \eqref{eq:h_coef}-\eqref{eq:f_coef}, the received complex baseband signal at the BS over RB$_r$ can be expressed as
\begin{align} \label{eq:sm_act_cluster}
    \resizebox{1\hsize}{!}{$
    y_r(\boldsymbol{\Delta})= \sqrt{p_\text{id}} \left(\sum_{\forall u \in \mathcal{C}_r} \sum_{\forall k, l, m}  \underbrace{\mathbf{g}_l^\mathrm{T} \left(  \mathbf{\Phi}_{k,l}^m \delta_{k,l}^m\right) \mathbf{h}_u^l  s_u }_{\text{UEs-RIS-BS}} + \underbrace{f_u  s_u }_{\substack{\text{UEs-BS}}}\right)  +  n,
    $}\raisetag{0.8\normalbaselineskip}
\end{align}
\begin{comment}
\begin{align} \label{eq:sm_pas_cluster}
    y_r^{pas} = \Bigg(\sum_{\forall u \in \mathcal{C}_r,\forall s \in \mathcal{S}}\underbrace{\mathbf{g}_b^\mathrm{T}\mathbf{\Phi}_b\mathbf{h}_u^i}_{\substack{\text{Reflected} \\ \text{link}}} + \sum_{\forall u \in \mathcal{C}_r}\underbrace{f_u}_{\substack{\text{Direct} \\ \text{link}}}\Bigg) \sqrt{p_u}s_u +  n,
\end{align}
\end{comment}
where $k\in \mathcal{R}$; $l\in \mathcal{B}$; $m\in \mathcal{U}$; $y_r(\boldsymbol{\Delta})$ is the received signal from members of $\mathcal{C}_r$ at RB$_r$; $s_u$ is the symbol transmitted by UE$_u \in \mathcal{C}_r$;  $n\sim\mathcal{CN}(0,\sigma^2)$ is the additive white Gaussian noise with variance $\sigma^2= N_0 W$; $N_0$ is the thermal noise power spectral density; $\mathbf{\Phi}_{k,l}^m = \mathrm{diag} \left( \begin{bmatrix} \phi_{k,l}^{m,1},\ldots, \phi_{k,l}^{m,n},\ldots,\phi_{k,l}^{m,N} \end{bmatrix} \right) \in \mathbb{C}^{N\times N}$ is the phase shift matrix of RIS$_l$ assigned to UE$_m \in \mathcal{C}_k$; and $\phi_{k,l}^{m,n}$ is the phase shift of the $n$th element of the RIS$_l$ configured as per UE$_m \in \mathcal{C}_k$. Assuming the CSI is acquired through accurate channel estimation methods \cite{Asmaa2022ICC} and $\delta_{k,l}^m=1$, the phase shifts of RIS$_b$ are configured as per the channel characteristics of UE$_m \in \mathcal{C}_k$
\begin{equation}
\label{eq:phasealign}
    \phi_{k,l}^{m,n} = e^{j\angle f_{m}} e^{-j\angle g_l^n}e^{-j\angle h_{m,l}^n},
\end{equation}
where $\angle \cdot$ denotes the phase of  complex numbers and element phases cancels the overall channel phases incurred from UE-RIS-BS cascaded channel and UE-BS direct channel. 

The SIC receiver at the BS iteratively decodes $y_r(\boldsymbol{\Delta})$ in the descending order of received signal power. That is, the cluster member with the strongest/weakest reception power is decoded first/last such that the strongest/weakest cluster member observes all/no intra-cluster interference from other cluster members. Correspondingly, $\widetilde{\mathcal{C}_r}$ denotes the cluster set when index set of the users is sorted in descending order according to the received signal power of the users. Accordingly, following from \eqref{eq:sm_act_cluster}, the SINR of $i$th ordered UE$_u \in \widetilde{\mathcal{C}_r}$ is given by
% \begingroup
% \setlength\abovedisplayskip{12pt}
% \setlength\belowdisplayskip{12pt}
% \begin{align} \label{eq:SINR_RIS}
%     \resizebox{1.001\hsize}{!}{$
%     \gamma_r^u(\boldsymbol{\Delta})=\frac{p_\text{id}\bigg(\sum_{\forall k, l, m} \lvert \mathbf{g}_l^\mathrm{T} \left(  \mathbf{\Phi}_{k,l}^m \delta_{k,l}^m\right) \mathbf{h}_u^l \rvert^2 + \lvert f_u \rvert^2 \bigg)} {p_\text{id}\bigg(\sum_{\substack{\forall z \in \widetilde{\mathcal{C}}_r[j>i],\\\forall k,l,m}}\lvert \mathbf{g}_l^\mathrm{T} \left(  \mathbf{\Phi}_{k,l}^m \delta_{k,l}^m\right) \mathbf{h}_z^l \rvert^2 + \lvert f_z \rvert^2 \bigg) + \sigma^2 },$}
% \end{align}
% \endgroup
\begin{equation} \label{eq:SINR_RIS}
    \resizebox{1.001\hsize}{!}{$
    \gamma_{r_\text{pas}}^u(\boldsymbol{\Delta})=\frac{p_\text{id}\bigg(\mathlarger{\sum}\limits_{\forall k, l, m} \Big\lvert \mathbf{g}_l^\mathrm{T} \left(  \mathbf{\Phi}_{k,l}^m \delta_{k,l}^m\right) \mathbf{h}_u^l \Big\rvert^2 + \lvert f_u \rvert^2 \bigg)} {p_\text{id}\left(\mathlarger{\sum}\limits_{\substack{\forall z \in \widetilde{\mathcal{C}}_r[j>i],\\\forall k,l,m}}\Big\lvert \mathbf{g}_l^\mathrm{T} \left(  \mathbf{\Phi}_{k,l}^m \delta_{k,l}^m\right) \mathbf{h}_z^l \Big\rvert^2 + \lvert f_z \rvert^2 \right) + \sigma^2 },$}\raisetag{0.5cm}
\end{equation}
where $j$ corresponds to the index of the users with lower received signal power than $i$th ordered user.
Even though the SINR expression in \eqref{eq:SINR_RIS} accounts for $s_u, \forall u \in \mathcal{C}_r$ reflected from RIS$_b, \forall b \in \mathcal{B}$, regardless of their assignment, the non-coherent signals reflected from RIS blocks assigned to other clusters have a negligible impact on the received signal from $\mathcal{C}_r$, $y_r(\boldsymbol{\Delta})$. Accordingly, \eqref{eq:SINR_RIS} can be simplified by merely considering the signal reflected from RIS$_b$ assigned to $\mathcal{C}_r$ as follows: 
% \begingroup
% \setlength\abovedisplayskip{12pt}
% \setlength\belowdisplayskip{12pt}
% \begin{align} \label{eq:SINR_RIS_simplified}
%     \resizebox{1.001\hsize}{!}{$
%     \gamma_r^u(\boldsymbol{\Delta})=\frac{p_\text{id}\bigg(\sum_{\forall m} \lvert \mathbf{g}_b^\mathrm{T} \left(  \mathbf{\Phi}_{r,b}^m \delta_{r,b}^m\right) \mathbf{h}_u^b \rvert^2 + \lvert f_u \rvert^2 \bigg)} {p_\text{id}\bigg(\sum_{\substack{\forall m,\\ \forall z \in \widetilde{\mathcal{C}}_r[j>i]}}\lvert \mathbf{g}_b^\mathrm{T} \left(  \mathbf{\Phi}_{r,b}^m \delta_{r,b}^m\right) \mathbf{h}_z^b \rvert^2 + \lvert f_z \rvert^2 \bigg) + \sigma^2 }.$}
% \end{align}
% \endgroup
\begin{equation} \label{eq:SINR_RIS_simplified}
    \resizebox{1\hsize}{!}{$\gamma_{r_\text{pas}}^u(\boldsymbol{\Delta})=\frac{p_\text{id}\bigg(\mathlarger{\sum}\limits_{\forall m}  \Big\lvert \mathbf{g}_b^\mathrm{T} \left(  \mathbf{\Phi}_{r,b}^m \delta_{r,b}^m\right) \mathbf{h}_u^b  \Big\rvert^2 + \lvert f_u \rvert^2 \bigg)} {p_\text{id}\left(\mathlarger{\sum}\limits_{\substack{\forall z \in \widetilde{\mathcal{C}}_r[j>i],\\\forall m }} \Big\lvert \mathbf{g}_b^\mathrm{T} \left(  \mathbf{\Phi}_{r,b}^m \delta_{r,b}^m\right) \mathbf{h}_z^b  \Big\rvert^2 + \lvert f_z \rvert^2 \right) + \sigma_\text{Rx}^2 }.$}\raisetag{0.4cm}
\end{equation}
\section{RIS Assisted Grant Free NOMA Scheme}
\label{sec:def}
In this section, we first provide a formal problem formulation, then outline the proposed solution methodology. 
\subsection{Problem Definition}
The joint user clustering and RIS assignment problem that maximizes overall network sum rate can be formulated as follows:
\begin{equation}
\label{prob:sum}
		\begin{aligned}
			& \hspace*{0pt} \vect{\mathrm{P}_1}: \underset{\boldsymbol{\mathcal{X}},\boldsymbol{\Delta}}{\max}
			& & \hspace*{1 pt} \sum_{ \forall r \in \mathcal{R}} \sum_{ \forall u\in \mathcal{U}}   W \log_2(1+ \chi_r^u \gamma_r^u(\boldsymbol{\Delta})) \\
			&\hspace{1.0cm}\text{s.t.} \\
			& \hspace*{0pt} \mbox{$\mathrm{C_1}$: }\hspace*{0.1pt} 
			&&   \gamma_r^u \geq 2^{q_u/B}-1 , \textbf{ } \forall u \in \mathcal{C}_r, \forall r \in \mathcal{R},\\ 
			&
			\hspace*{0 pt}\mbox{$\mathrm{C_2}$: } & &  \sum_{r \in \mathcal{R}} \chi_r^u = 1, \textbf{ }  \forall u \in \mathcal{U}, \\
			&
			\hspace*{0 pt}\mbox{$\mathrm{C_3}$: } & & \sum_{\forall u \in \mathcal{U}} \chi_r^u \leq K, \textbf{ }  \textbf{ }  \forall r \in \mathcal{R}, \hspace{0.75 cm}  \\
			&
			\hspace*{0 pt}\mbox{$\mathrm{C_4}$: } & & \delta_{r,b}^u \leq \chi_r^u , \textbf{ }  \textbf{ }  \forall r \in \mathcal{R}, \forall b \in \mathcal{B}, \forall u \in \mathcal{U}, \hspace{0.75 cm}  \\
			&
			\hspace*{0 pt}\mbox{$\mathrm{C_5}$: } & &  \sum_{\forall u \in \mathcal{U}} \sum_{\forall b \in \mathcal{B}} \chi_r^u \delta_{r,b}^u \leq 1, \textbf{ }  \textbf{ }  \forall r \in \mathcal{R}, \hspace{0.75 cm}  \\	
			&	
			\hspace*{0 pt}\mbox{$\mathrm{C_6}$: } & &  \sum_{\substack{\forall u \in \mathcal{U}}} \sum_{\substack{\forall r \in \mathcal{R}}} \sum_{\substack{\forall b \in \mathcal{B}}} \delta_{r,b}^u \leq B, \textbf{ }  \textbf{ } \hspace{0.75 cm}  \\	&		
			\hspace*{0 pt}\mbox{$\mathrm{C_7}$: } & & \chi_r^u \in \{0,1\}, \delta_{r,b}^u \in \{0,1\}, \forall r, \forall u, \forall b.
		\end{aligned},
\end{equation}
Here, $\gamma_{r}^u$ is the signal-to-noise-plus-interference-ratio (SINR) of UE$_u \in \mathcal{C}_r$, $\boldsymbol{\mathcal{X}} \in \{0,1\}^{U \times R}$ is the binary UE clustering matrix with entries $\chi_r^u$. In $\vect{P_1}$, $\mathrm{C_1}$ ensures that UE$_u$ is satisfied with the quality of service demand $q_u$ [bps] $\forall u \in \mathcal{C}_r, \forall r \in \mathcal{R}$, $\mathrm{C_2}$ assures each UE is admitted to a cluster, $\mathrm{C_3}$ limits the cluster size by $K=\lceil U/R \rceil$, $\mathrm{C_4}$ states that RIS$_b$ cannot be configured as per UE$_u$ if it is not a member of $\mathcal{C}_r$, $\mathrm{C_5}$ guarantees each cluster is assisted by at most one RIS block, $\mathrm{C_6}$ limits the total RIS assignments by the total number of RIS blocks, and $\mathrm{C_7}$ specifies the domain and bounds on optimization variables. $\vect{\mathrm{P}_\mathrm{1}}$ is a mixed integer nonlinear programming (MINLP) problem, which is non-convex due to the interference terms in the SINR expression defined in the next section. Finding an optimal solution to this NP-Hard problem is computationally prohibitive even for a moderate size of network. This necessitates a heuristic solution for real-life implementation, which is discussed next. 
\subsection{Proposed Iterative User Clustering and RIS Assignment}
\label{sec:Iterative}
The proposed RIS-assisted GF-NOMA scheme allows UEs to access RBs at any time, requiring neither grant acquisition from the BS nor power control at the UE side. Instead, the power disparity requirement of GB PD-NOMA is satisfied by user clustering and RIS assignment through a three-level implicit power control: 

\begin{enumerate}
    \item The first level of power control is obtained by pairing users with different channel gains on the same cluster/RB. Even if UEs operate at identical transmit power, the channel gain disparity introduces a reception power disparity to achieve a higher PD-NOMA gain.   
    \item The second level of power control is obtained by assigning a proper RIS block to a cluster. It is worth noting that RIS improves overall communication performance if it is deployed near the UE or the BS \cite{kilinc2021physical}. Therefore, RIS block assignment plays a crucial role in increasing the power reception disparity for an improved cluster sum rate.
    \item The final level of power control is obtained by selecting the cluster member according to which assigned RIS is configured. Since reception power levels determine the SIC decoding order, the cluster member selection is also critical to improve the overall cluster sum rate.  
\end{enumerate} 
Before delving into the details of the proposed RIS-assisted GF-NOMA scheme, it is worth reminding that the GB optimal PD-NOMA scheme requires UEs with stronger channels to transmit at the higher powers to maximize the UL-NOMA cluster sum rate \cite{celik2022gfnoma}. However, this approach is not fair in energy consumption and may result in reduced network lifetime, which is paramount for low-power machine-type communications. Alternatively, the proposed approach improves network lifetime by requiring all UEs to transmit at identical power fairly and constitutes the required power disparity through iterative user clustering and RIS assignment summarized in Algorithm 1. Following cluster initialization, the proposed iterative approach admits a single UE to each cluster at each iteration until all UEs are admitted to a cluster. To this aim, the admissions are determined based on a three-dimensional axial assignment (3D-AA) that yields the maximum network sum rate. The 3D-AA is known to be an NP-Hard problem, whose matching theory-based approximate solutions can obtain results for a square cost matrix with complexity $\mathcal{O}\left(3MV^3\right)$ where $M$ is the number of relaxations and $V$ is the largest dimension of the cost matrix \cite{3Dcomplexity}.
\begin{figure}[t!]
    \rule{\columnwidth}{0.85pt}
    \textbf{Algorithm 1 : Joint UE Clustering and RIS Assignment}
    \rule{\columnwidth}{0.5pt}
    \vspace{-3pt}
	\fontsize{8pt}{8pt}\selectfont
		%\caption{\textbf{: Joint UE Clustering and RIS Assignment}}
		\label{alg:joint}
		\begin{algorithmic}[1]
		
			\renewcommand{\algorithmicrequire}{\textbf{Input:}}
			\renewcommand{\algorithmicensure}{\textbf{Output:}}
			
			\State \hspace{-5pt} \textbf{Input: } $\mathcal{R}$, $\mathcal{U}$, $\mathcal{B}$, P
			
			\State $\mathbf{h}_u^b/\mathbf{g}_b/f_u \gets $ \textit{Acquire CSI from SRS sent by UE$_u, \forall u \in \mathcal{U}, \forall b \in \mathcal{B}$} \label{line:rss}

			\State $C \gets R$  \hfill \textit{// Determine number of clusters} \label{line: C}	
			
			\State $K\gets \left \lceil \frac{U}{R} \right \rceil$ \hfill  \textit{//  Determine maximum cluster size} \label{line:Up}
			
		    \State $\tilde{\mathcal{U}} \gets$ \textsc{SortDescend}($\mathfrak{Rss}, \forall u$)   \hfill \textit{// UE ordering based on RSS of SRS} \label{line:sort_user}
		    
			\State ${\mathcal{C}}_r \gets \tilde{\mathcal{U}}\{r\},  r \in [1,\ldots, R]$  \hfill \textit{// Initialize clusters} \label{line:init_C}
			
			\State $\chi_r^u \gets 1, u \in {\mathcal{C}}_r$  \hfill \textit{//Initialize UE clustering matrix} \label{line:init_chi}
			
			\For{$k\!=\!1\!:\!K\!-\!1$}  \textit{// Iterative UE admission and RIS Assignment starts}
			\label{line:adm_begin}
			
    			\State ${\mathcal{A}_k} \gets \tilde{\mathcal{U}} - \bigcup_{r=1}^R \mathcal{C}_r $ \hfill \textit{// Initialize admission awaiting UEs}\label{line:An}
        
    			\State $  \vect{Q_k}, \vect{I_k} \gets $ \textsc{Cost Matrix}($\vect{C}_r,\mathcal{A}_k,\mathcal{B}$) \label{line:costmatrix} \hfill \textit{// Generating cost matrix}
    			
				\State $\accentset{}{\boldsymbol{Y}_k} \gets$ \textsc{3D-Axial Assignment}($\vect{Q_k}, \mathcal{A}_k$)   \label{Line:3DAssignment}
    			
     			\State $\mathcal{C}_r \gets \mathcal{C}_r \bigcup  \mathcal{A}_{k} \{u\}, y_{r,b}^u=1, \forall (r, u, b)$ \hfill \textit{// Update clusters}\label{line:update_C}  
     			
     			\State $\chi_r^u \gets 1, u \in {\mathcal{C}}_r$  \hfill \textit{// Update UE clustering matrix} \label{line:update_chi}
     			
     			\State $\delta_{r,b}^{u'} \gets 1$ iff $u'=\iota_{r,b}^u$, $u\in\mathcal{C}_r, \forall (r,b) $   \textit{//Update RIS assignments} \label{line:update_delta}
			\EndFor \label{line:adm_end}
			\\
			\Return $\boldsymbol{\mathcal{X}}$, $\boldsymbol{\Delta}$
			\vspace{2pt}
			\hrule 
			\vspace{2pt}
			\hrule 
			\vspace{2pt}
			
			\Procedure{Cost Matrix}{$\vect{C}_r,\mathcal{A}_k,\mathcal{B}$} \label{line:costbegin}
			\State $\vect{Q},\vect{I} \gets \vect{0}^{R \times A \times B}$ \hfill \textit{// Initialize $\vect{Q}$ and $\vect{I}$ to matrix of zeros}
			\For{$r=1:\mathcal{R}$} \label{line:fori}
			\For{$b=1:B$} \label{line:fork}
			\For{$u=1:A$}  \label{line:forj}
    			\State $\mathcal{T}_r \gets \mathcal{C}_r \bigcup \mathcal{A}_k \{u\} $ \hfill \textit{// Temp. admit $u^{th}$ UE of $\mathcal{A}_k$ to $\mathcal{C}_r$} \label{Line:Tr}
    			
    			\For{$j\in T_r$} \label{Line:iterue}
 		        \State $\mathbf{\Phi}_{r,b}^j \gets$ Align RIS$_b$ to UE$_{\mathcal{T}_r\{j\}}$ as per \eqref{eq:phasealign} \label{Line:phasealign}
 		        \State $\gamma_r^j \gets$ Obtain SINR as per \eqref{eq:SINR_RIS_simplified} \label{Line:SINR}
 		        \State $\eta_j^\star \gets \sum_{j \in \mathcal{T}_r} W \log_2(1+\gamma_r^j)$  \label{line:ul_sumrate}
 			 \EndFor \label{Line:iterueend}
                \State $q_{r,b}^u \gets \max(\eta_j^\star, j \in \mathcal{T}_r)$ \hfill \textit{// Update cost matrix entries} \label{Line:maxq}
 			 \State $\iota_{r,b}^u \gets {\argmax}(\eta_j^\star, j \in \mathcal{T}_r)$ \hfill \textit{// Update $\vect{I}$ entries} \label{Line:maxq_ind}
			\EndFor
			\EndFor
			\EndFor
			\\
			\Return $\boldsymbol{Q}, \boldsymbol{I}$
			\EndProcedure \label{line:cost_end}
			\vspace{2pt}
			\hrule 
			\vspace{2pt}
			\hrule 
			\vspace{2pt}
			\Procedure{3D-Axial Assignment}{$\vect{Q}, \mathcal{A}_k$} \label{line:3Dbegin}
            \State $\accentset{\star}{\vect{Y}} \gets \underset{\vect{Y}}{\max} \: \:  \sum_{r \in \mathcal{R}} \sum_{b \in \mathcal{B}} \sum_{u \in \mathcal{U}} q_{r,b}^u y_{r,b}^u $ \\ 
			\hspace{35pt} $ \text{s.t. } \hspace{5pt} \sum_{r \in \mathcal{R}} \sum_{b \in \mathcal{B}} y_{r,b}^u = 1, \forall u \in \mathcal{A}_k$ \label{line:3dconst1} \\
		   \hspace{54pt} $ \sum_{r \in \mathcal{R}} \sum_{u \in \mathcal{U}} y_{r,b}^u = 1, \forall b \in \mathcal{B}$ \label{line:3dconst2} \\
		   \hspace{54pt} $ \sum_{b \in \mathcal{R}} \sum_{u \in \mathcal{U}} y_{r,b}^u = 1, \forall r \in \mathcal{R}$ \label{line:3dconst3} \\
			\Return $\accentset{\star}{\vect{Y}}$
			\EndProcedure \label{line:3Dend}
		\end{algorithmic}
	\rule{\columnwidth}{0.5pt}
\end{figure}

Algorithm 1 starts with receiving sounding reference signals (SRS), which is a Zadoff-Chu sequence transmitted by each UE separately from the Physical UL shared channel (PUSCH) and physical uplink control channel (PUCCH). UEs can transmit SRS on any subcarriers in the last symbol of an uplink subframe regardless of subcarriers assignments. We especially focus on time-division duplexing (TDD) mode for the sake of channel reciprocity such that SRS may also be sent in the last two symbols of the special subframe if UL pilot time slot (UpPTS) is configured to be in the long format. The SRS is imperative to estimate cascaded and direct channels \cite{Asmaa2022ICC} to determine RIS assignment and configure phase shift matrices accordingly. Notice that the CSI acquisition is an inherent part any communication standard and not specific to the proposed approach. After the cluster size is determined based on the available number of RBs in Line \ref{line: C}, the maximum cluster size is determined in Line \ref{line:Up} based on the UE density. For the sake of cluster initialization, the users are sorted in the descending order of received signal strength of SRS signals in Line \ref{line:sort_user}. Then, the clusters are initialized by admitting the first $R$ users into $C$ clusters in Line \ref{line:init_C}. Then, the iterative user admission and RIS assignment start in Line \ref{line:adm_begin} where one more user is admitted to each cluster in each iteration. Therefore, the first line of for loop, Line  \ref{line:An}, updates the set of UEs awaiting cluster admission, which is denoted by $\mathcal{A}_k$. 

To execute 3D-AA, Line \ref{line:costmatrix} calls cost matrix generation procedure given between Line \ref{line:costbegin} and Line \ref{line:cost_end}, where cost matrix $\boldsymbol{Q} \in \mathbb{R}^{R \times A \times B}$ and temporary RIS alignment matrix $\boldsymbol{I} \in \mathbb{N}^{R \times A \times B}$ are formed over three nested for loops. In Line \ref{Line:Tr}, the $u$-th element of admission awaiting user set is temporarily admitted to $\mathcal{C}_r$, where $A$ denotes the length of the set $\mathcal{A}_k$. Then, the most inner loop between Line \ref{Line:iterue} and Line \ref{Line:iterueend} determines the cluster member giving the maximum cluster sum rate if RIS$_b$ is aligned as per its channel conditions. To this aim, Line \ref{Line:phasealign} adjusts the phase shift matrix of RIS$_b$ to UE$_{\mathcal{T}_r\{j\}}$, Line \ref{Line:SINR} calculates the cluster members' SINR based on  adjusted the phase shift matrix as per \eqref{eq:SINR_RIS_simplified}, and finally Line \ref{line:ul_sumrate} records the cluster sum rate of $\mathcal{T}_r$ if RIS$_b$ is aligned to its $j$-th member, $\eta_j^\star$. Accordingly, the cost matrix element $q_{r,u}^b$ is updated in Line \ref{Line:maxq} based on the user cluster member giving the highest possible cluster rate if the phase shift matrix of RIS$_b$ is aligned to itself, whose index matrix (representing the user index) is also stored in a temporary RIS alignment matrix in Line \ref{Line:maxq_ind}. 
%Accordingly, the cost matrix element $q_{r,b}^u$ is updated in Line \ref{Line:maxq} based on the cluster member $\mathcal{C}_r\{\iota_{r,b}^u\}$ giving the highest possible cluster rate if the phase shift matrix of RIS$_b$ is aligned to itself, whose index in the cluster $\iota_{r,b}^u$, which is also stored in a temporary RIS alignment matrix in Line \ref{Line:maxq_ind}.

Following the generation of the cost-matrix, Line \ref{Line:3DAssignment} calls \textsc{3D-Axial Assignment} to find joint user clustering and RIS assignment at $k$-th iteration. For instance, when 3D-AA returns $y_{3,4}^5=1$, it means that: 1) UE$_{\mathcal{A}_k\{5\}}$ is admitted to $\mathcal{C}_3$, 2) RIS$_4$ is assigned to $\mathcal{C}_3$, and 3) phase shift matrix of RIS$_4$ is aligned to assisted user UE$_{u'}$ whose user index stored in $u'=\iota_{3,4}^5$. Thereafter, the cluster members are updated as per the outcome of 3D-AA in Line \ref{line:update_C}, which is followed by update of user clustering matrix $\boldsymbol{\mathcal{X}}$ and RIS assignment matrix $\boldsymbol{\Delta}$ in Line \ref{line:update_chi} and \ref{line:update_delta}, respectively. The final user clustering and RIS assignment matrices are returned at the end of $(K-1)$-th iteration. Notice in Line \ref{line:update_delta} that the phase alignment is obtained from the RIS alignment matrix $\boldsymbol{I}$ returned by the \textsc{Cost Matrix} procedure. 

The time complexity of the proposed solution is mainly driven by \textsc{Cost Matrix} and \textsc{3D-Axial Assignment} procedures. While the complexity of cost matrix computation is $\mathcal{O}\left(\sum_{k=1}^{K-1} R A_k B\right)$ and the time complexity of the 3D-AA is $\mathcal{O}\left(3 M \sum_{k=1}^{K-1} (V_k)^3\right)$, where $V_k=\max(R,A_k,B)$. Since the latter is more dominant, the overall complexity can be approximated as $\sim \mathcal{O}\left(3 M \sum_{k=1}^{K-1} (V_k)^3\right)$.

\section{Numerical Results}
\label{sec:numresult}
% \subsection{Benchmark NOMA Schemes}

% In this section, three NOMA schemes are presented as benchmarks for the RIS-Assisted GF-NOMA scheme. 

In this section, we provide numerical results for the proposed scheme and compare its performance with three benchmark schemes for different system parameters. Achievable rates for different schemes are presented by computer simulations. We consider the systems with parameters $M=25$, $G=1$, $U=75$, $R=25$, $C=25$, $p_\text{id}=21$ dBm, $\mathrm{W}=180$ kHz, $q_u=10^5$ bps, $N=256$, $f_c=5$ GHz, $D_\text{in}=15$ m, $D_\text{out}=50$, and $D=250$ m, $N_0=-174$ dBm, $P_\text{max}=23$ dBm and $P_\text{min}=-40$ dBm, unless stated otherwise. Although we consider the SINR expression in \eqref{eq:SINR_RIS_simplified} for RIS assignment in the algorithm, we consider the general SINR expression in \eqref{eq:SINR_RIS} while obtaining the numerical results, so that the reflections from non-coherent RIS blocks are considered as well. Achievable rate simulations are obtained by MATLAB software and averaged over $10^6$ network realizations. The proposed solution is compared with the following benchmark schemes \cite{celik2022gfnoma}:
\begin{enumerate}
    \item 
    The optimal (OPT) PD-NOMA where UEs compute optimal power control based on the readily available CSI, and the maximum and minimum transmit power of UEs are $P_\text{max}$ and $P_\text{min}$, respectively. Throughout the simulations, optimal power weights are obtained by geometric programming solver of the CVX disciplinized convex optimization toolbox \cite{cvx}.
    \item 
    Multi-level GF-NOMA (MGF-NOMA) scheme partitions the set of UEs into $R=C$ groups based on their channel gains and divides power control range ($[p_\textit{max}=23, p_\textit{min}=-40]$ dBm) into $R=C$ levels. MGF-NOMA requires UE partitions with higher channel gain to transmit at higher power levels. For instance, the UE partition consisting of UEs with the highest/lowest channel gains are required to transmit at the highest/lowest power levels. 
    \item 
    Single level GF-NOMA (SGF-NOMA) scheme requires all UEs transmit at an identical transmit power, $p_\text{id}$.
\end{enumerate}
We refer interested readers to \cite{celik2022gfnoma} for a more detailed explanation of the benchmark schemes. For the sake of a fair comparison, the proposed schemes uses $p_\text{id}=21$ dBm for all UEs, which is the average transmit power of UEs for OPT PD-NOMA. Noting that benchmark schemes do not benefit from RIS, the main difference between compared schemes is the underlying power control approach.
% \subsubsection{Optimal Grant-Based PD-NOMA}

% In the optimal GB PD-NOMA system, UEs are assumed to have perfect CSI of the cluster members and optimize their transmit power to maximize the cluster sum rate. The optimal power levels can be obtained through analytical or numerical methods \cite{celikclustering}. 

% \subsubsection{Multi-Level Grant-Free (MGF) PD-NOMA}
% In MGF-NOMA, UEs are partitioned into $K$ subsets based on their channel gain. Likewise, the power control range of UEs, $[23,-40]$ dBm, is also divided into $K$ linearly spaced levels. The power levels are assigned to each partition proportional to their channel gains such that the UE subset with the strongest and weakest channels transmit at the highest (23 dBm) and lowest (-40 dBm) power levels.

% \subsubsection{Single-Level Grant-Free (SGF) PD-NOMA}
% In the SGF-NOMA, all users transmit at the identical $P$ transmit power. The required power difference between users is created by exploiting the difference in channel gains.

% Although the benchmark schemes differ in power control, they follow a common user clustering scheme where clusters are initialized with the $R$ highest channel gain UEs. Then, an iterative 2D assignment approach is leveraged to admit a UE to clusters at each iteration until all UEs are admitted to $C$ clusters. 

\begin{figure}[t!] \centering \subfloat{\includegraphics[width=\columnwidth]{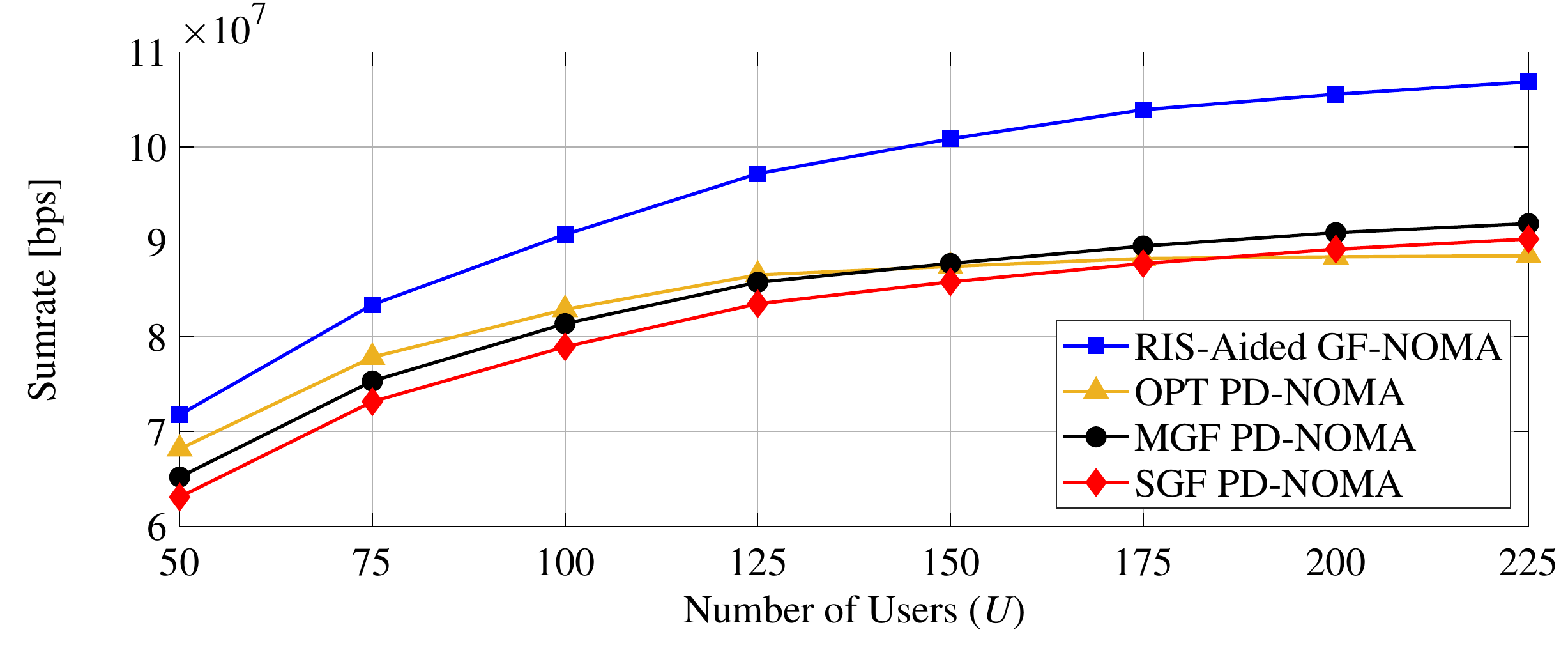}}
\vspace{-0.05cm}
 
\subfloat{\includegraphics[width=\columnwidth]{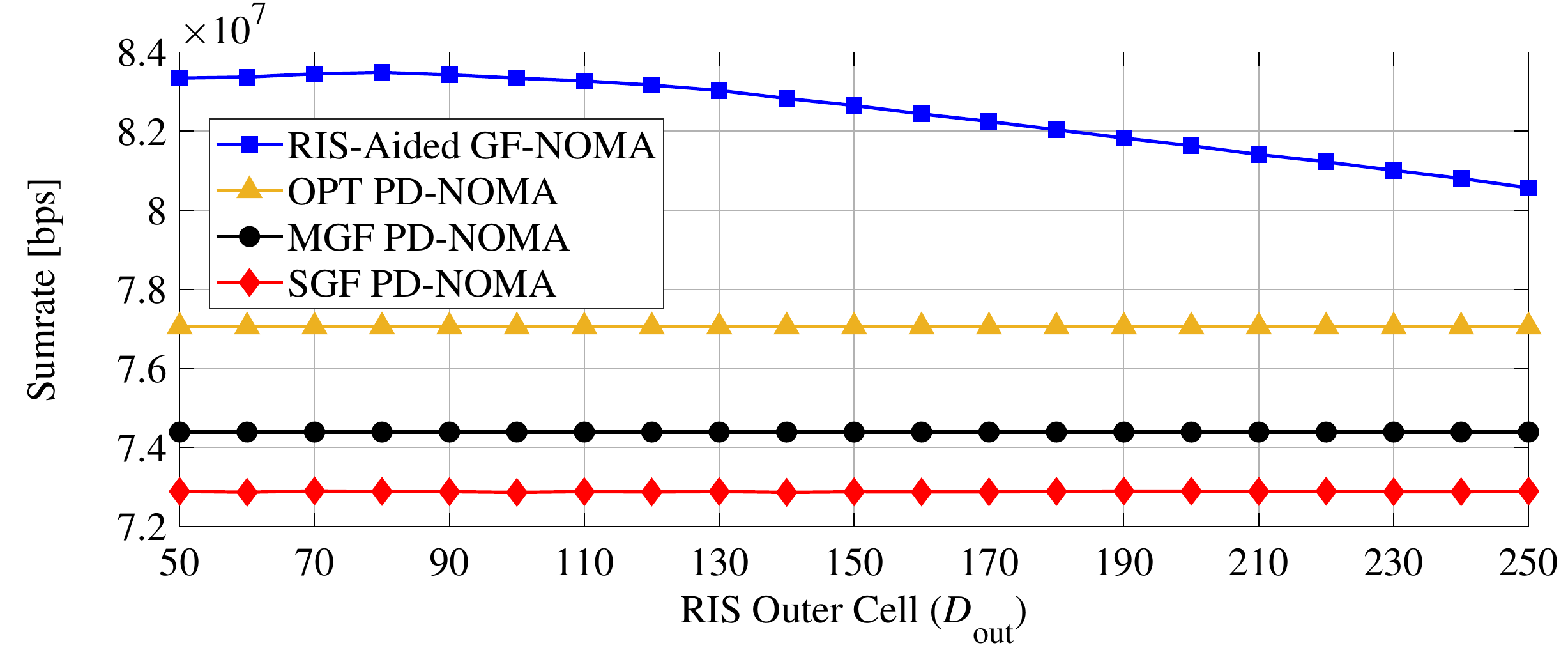}}
 
\subfloat{\includegraphics[width=\columnwidth]{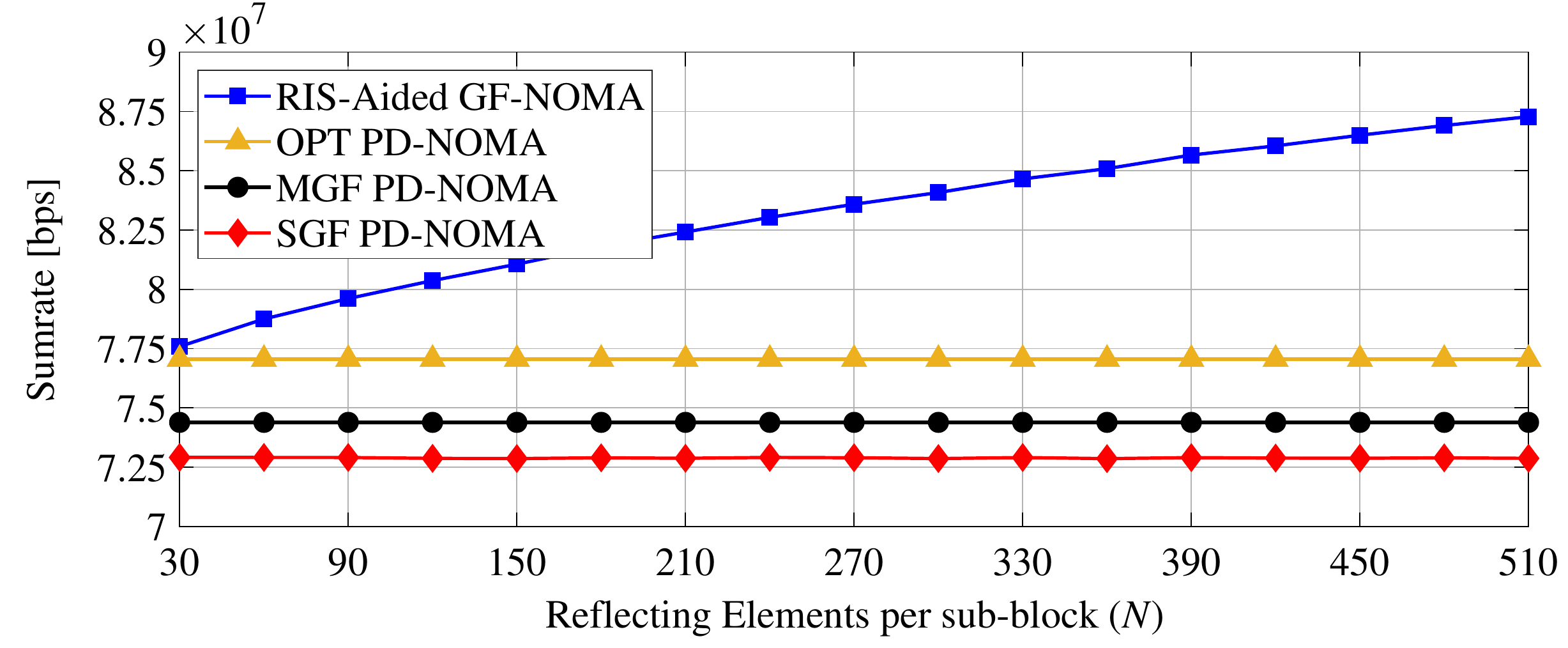}}
\caption{Network sum rate for different metrics: (a) $U$, (b) $D_\text{out}$, and (c) $N$.} \label{fig:numresults}\vspace*{-0.55cm} 
\end{figure}
The number of users in a cluster is an important metric that affects the network performance. Fig. \ref{fig:numresults}(a) exhibits the achievable rate of the network for varying $U$. As seen in Fig. \ref{fig:numresults}(a), the sum rate of the proposed scheme increases with $U$ because the number of users operating in a single resource block increases, resulting in more efficient spectrum usage. The proposed scheme respectively performs $15\%$, $15\%$ and $18\%$ better than OPT PD-NOMA, MGF-NOMA and SGF-NOMA schemes when $U=150$. We note that the increase in the cluster size does not have a considerable negative effect on the network performance. The proposed scheme performs better on denser networks, such as where $U=150$ and the number of users per cluster is $6$ and outperforms the benchmark schemes. The performance of the benchmark schemes starts to be saturated after $U=150$ while the proposed scheme continues to show an upwards trend.  %8 12 15 dout=80  ...  5 8 11 dout = 250  %%%% %8 12 14 N=240 ... 13 17 20 N = 510

Similar to the number of users in a cluster, the locations of the RISs can also affect the system performance. It is a well-known fact that a basic RIS-assisted system exhibits its maximum performance when the RIS is located near the terminals of the system. Similarly, the proposed design also gives the best performance when the RISs are close to the BS or users. Fig. \ref{fig:numresults}(b) shows the sum rate of the network for varying $D_\text{out}$. The results show that centralized RIS deployment performs better than distributed RIS deployment scenarios because we cannot guarantee that the RISs will be placed close to the users since users are randomly distributed as well. The proposed scheme performs $8\%$, $12\%$ and $15\%$ better than OPT PD-NOMA, MGF-NOMA and SGF-NOMA schemes, respectively, when $D_\text{out}=80$ m. As $D_\text{out}$ increases, the sum rate decreases because the RISs gradually become distant from the BS and most of the users. Nevertheless, the proposed scheme still performs $5\%$, $8\%$ and $11\%$ better than OPT PD-NOMA, MGF-NOMA and SGF-NOMA scheme, respectively, when $D_\text{out}=250$ m. We previously mentioned that an RIS could assist the other users in different clusters, even if its phases are aligned for the dedicated user. This condition also helps further improvement in the sum rate when the RIS placed close to the BS.

RIS size has a considerable impact on the achievable rate of RIS-assisted systems. As shown in Fig. \ref{fig:numresults}(c), the proposed system has nearly the same sum rate with OPT PD-NOMA scheme when $N=30$, but still performs better than SGF-NOMA scheme by 7\%. When $N=510$, the proposed scheme performs $13/17/20\%$ better than OPT PD-NOMA/MGF-NOMA/SGF-NOMA scheme. The increase in $N$ further increases the performance of the proposed scheme while not affecting the performance of the benchmark schemes since they are not RIS assisted.\vspace{-1pt}
\section{Conclusion}
\label{sec:conc}
In this paper, an iterative clustering algorithm with a 3D assignment approach has been proposed for an RIS-assisted GF-NOMA system. Through the proposed clustering approach, the network has been found to be able to leverage from the RIS assistance to create received power disparity among NOMA users without the need for complex power control. Moreover, assisting the system with RISs results in better SIC due to the power disparity introduced from the presence of the RISs, leading to improved achievable rates compared with the conventional GF-NOMA with no RIS assistance. Future works can be considered for new schemes that use active RISs instead of the passive ones and assist not only one user in the cluster but all of them, which can come up with some complex optimization problems.\vspace{-2pt}

% trigger a \newpage just before the given reference
% number - used to balance the columns on the last page
% adjust value as needed - may need to be readjusted if
% the document is modified later
%\IEEEtriggeratref{8}
% The "triggered" command can be changed if desired:
%\IEEEtriggercmd{\enlargethispage{-5in}}

% references section

% can use a bibliography generated by BibTeX as a .bbl file
% BibTeX documentation can be easily obtained at:
% http://mirror.ctan.org/biblio/bibtex/contrib/doc/
% The IEEEtran BibTeX style support page is at:
% http://www.michaelshell.org/tex/ieeetran/bibtex/
%\bibliographystyle{IEEEtran}
% argument is your BibTeX string definitions and bibliography database(s)
%\bibliography{IEEEabrv,../bib/paper}
%
% <OR> manually copy in the resultant .bbl file
% set second argument of \begin to the number of references
% (used to reserve space for the reference number labels box)
%\newpage
\bibliographystyle{IEEEtran}
\bibliography{ref}

\flushend

% that's all folks
\end{document}